\documentclass[aps,prb,twocolumn,groupedaddress,floatfix]{revtex4}
\usepackage{graphics,graphicx}
\usepackage{color}
\usepackage{amssymb,amsmath}
\usepackage{bm}
\newcommand{\req}[1]{Eq.~(\ref{#1})}
\newcommand{\reqs}[1]{Eqs.~(\ref{#1})}
\newcommand{\rref}[1]{(\ref{#1})}

\newcommand{\vare}{\omega}

\newcommand{\be}{\begin{equation}}
\newcommand{\ee}{\end{equation}}
\newcommand{\bea}{\begin{eqnarray}}
\newcommand{\eea}{\end{eqnarray}}
\begin{document}
\unitlength = 1mm
%~~~~~~~~~~~~~~~~~~~~~~~~~~~~~~~~~~~~~~~~~~~~~~~~~
\title{Coexistence between superconducting and spin density wave states in iron-based superconductors:
Ginzburg-Landau analysis}
\author{M.~G.~Vavilov$^1$, A.~V.~Chubukov$^1$, and  A.~B.~Vorontsov$^2$ }
\affiliation{$^1$~Department of Physics,
             University of Wisconsin, Madison, Wisconsin 53706, USA\\
$^2$~Department of Physics, Montana State University, Bozeman, MT, 59717, USA}

\date{December 17, 2009}

%~~~~~~~~~~~~~~~~~~~~~~~~~~~~~~~~~~~~~~~~~~~~~~~~~~~~~~~~~~~~~~~~~~~~~~~~~~~~~
\begin{abstract}
We consider the interplay between superconducting (SC) and commensurate spin-density-wave (SDW) orders in iron-pnictides by analyzing a multiple order Ginzburg-Landau free energy.  We are particularly interested in whether the doping-induced transition between the two states is first order, or the two pure phases are separated by an intermediate phase with coexisting  SC and SDW orders.  For perfect nesting, the two orders do not coexist,  because SDW order, which comes first, gaps the full Fermi surface leaving no space for SC to develop.  When nesting is not perfect due to either ellipticity of electron bands or doping-induced difference in chemical potentials for holes and electrons, SDW order still leaves modified Fermi surfaces  for not too strong SDW magnetism and the SC order may develop.  We show that the two orders coexist only when certain relations between ellipticity and doping are met. In particular, in a compensated metal, ellipticity alone is not sufficient for coexistence of the two orders.  
\end{abstract}
%~~~~~~~~~~~~~~~~~~~~~~~~~~~~~~~~~~~~~~~~~~~~~~~~~~~~~~~~~~~~~~~~~~~~~~~~~~~~~
\maketitle
%~~~~~~~~~~~~~~~~~~~~~~~~~~~~~~~~~~~~~~~~~~~~~~~~~~~~~~~~~~~~~~~~~~~~~~~~~~~~~

\section{Introduction.}

In the phase diagram of recently discovered iron-based pnictide materials 
superconducting (SC) and spin density wave (SDW) states 
 are close neighbors.~\cite{Kamihara08}  The interplay between these two orders  has been 
the focus of numerous experimental and theoretical studies. Superconductivity
and magnetic ordering are normally mutually exclusive states of electronic
systems, and a first-order transition between SC and SDW
orders has been reported in some pnictides, e.g. for LaO$_{1-x}$F$_x$FeAs.\cite{Luetkens09}
 However, recent nuclear magnetic resonance (NMR),~\cite{Laplace09} specific heat, susceptibility, Hall
coefficient,~\cite{Chu09,Rotter08} and neutron scattering experiments~\cite{nature} on
Ba(Fe$_{1-x}$Co$_x$)$_2$As$_2$ indicate that SDW and SC phases coexist over some doping range. 

Electronic spectrum of the pnictides results in two hole pockets centered at
$(0,0)$ and two electron pockets centered at $(0,\pi)$ and $(\pi,0)$ in
the unfolded Brillouin zone (BZ).~\cite{ARPES-Liu08,ARPES-Evtushinsky09,ARPES-Hsieh08,Ding08-1,Zabolotnyy09,Coldea08,Coldea09,SinghDu-PRL100,phonons,mazin:057003}
Multiple Fermi
surfaces (FSs) create a number of different possibilities\cite{chubukov08,chubukov09} for
electron ordering in the form of SDW, charge density wave (CDW) states, and a
superconducting state with extended $s-$wave symmetry in which the gaps on the
hole and electron FSs are of different signs (an $s^{+-}$
state).~\cite{mazin:057003,Kuroki08,Barzykin08,Maier-PRB79,CVV-PRB09,Thomale09,Cvetkovic09,Stanev08}
When one order develops, electron states are
reorganized and either favor or hinder the development of other orders.
Specifically, in case of a compensated metal and a perfect nesting 
 (all FSs are cylinders of equal radius)
an SDW order comes first 
 and completely gaps all Fermi surfaces preventing the system from developing
 SC or CDW orders.  Situation is different, however, if electron and hole
 pockets are not ideal circles and intersect at certain points when the two FSs
 are centered at the same momentum. 
 This is what is actually observed in the pnictides: hole FSs have circular cross sections, while  
 electron FSs have elliptic ones. 
 In this case the SDW order modifies the Fermi surfaces 
 and leaves ungapped electronic states that may develop 
 superconducting instability.~\cite{parker} 
 Another possibility for coexistence is formation of 
 an SDW order with incommensurate wavelength on circular 
 hole and electron FSs of different radii. 
 Again, strong nesting of electronic states occurs only on 
 a small part of the FS, leaving other parts available for 
 the SC pairing.~\cite{anton} 
The persistence of the (modified) FS into the SDW phase, however, only implies that a superconducting order {\it may} develop within the SDW phase.~\cite{parker,anton} 
The relative thermodynamic stability of different 
phases (SC, SDW, SC+SDW) can only be determined from the analysis of the free energy.

In the present manuscript we report analytical study of the nature of the phase
transition between SC and SDW states in iron-pnictide materials. We demonstrate
that  SDW   and extended $s^{+-}$ SC orders 
  may coexist, but
  only in a situation  when both ellipticity {\it and} a difference $\delta \mu$ in electron and hole chemical potentials 
are present. 
If either ellipticity or $\delta \mu$  are small,
  SDW and SC states are separated by a first order phase transition. In
 particular, if the system remains a compensated metal, the transition is first
 order, even when electron bands are elliptical.  
 
 We also discuss the  $s^{++}$  order parameter 
 (ordinary s-wave). The authors of Ref.~\onlinecite{nature} argued that for this
 order parameter coexistence is impossible for any
  FS geometry.  The argument was based on the observation that the transition is
 strongly first order 
  at a perfect nesting,
  and on numerical results for a finite ellipticity and a finite $\delta \mu$.
 We analyzed this issue analytically and confirmed that the transition between
 SDW and $s^{++}$ SC states is indeed first order.

We follow earlier works~\cite{anton,parker,nature} and consider a simplified
model with  one circular hole and one elliptical electron FS, separated by,
e.g., $(0,\pi)$ in the unfolded BZ, or $(\pi,\pi)$ in the folded zone. This
momentum is also the ordering momentum of the commensurate SDW state.\cite{de08,Klauss08}
The
inclusion of another hole and electron FSs does not affect either
superconductivity,
\cite{Maier-PRB79,CVV-PRB09,Thomale09,Cvetkovic09,dhlee-PRL102,Platt09}
or  magnetic
order.\cite{Lorenzana08,Brydon09,johannes:220510,mazin09,Eremin09}
We will also assume that SDW order at a finite doping  remains
commensurate. This is true if SDW and SC instabilities, when taken
separately, occur not far from each other (we set the condition below). If this
condition is not met, SDW order becomes incommensurate even before SC develops,
and the interplay between SDW and superconductivity has to be re-evaluated. For
circular FSs this has been done in Refs.~\onlinecite{anton,Cvetkovic09}, and the result
is that an incommensurate SDW order and an $s^{+-}$ superconductivity do
coexist, at least near $T_s$.      

\section{General form of the free energy}

We first present a general analysis of the free energy for a system
characterized by two scalar order parameters, which we denote as $\Delta$ for an SC order
and
$M$ for an SDW order.   We
assume that $\Delta$  has the same momentum-independent magnitude, 
 but may have either equal signs ($s^{++}$ state)
or opposite signs ($s^{+-}$ state)
on the hole and electron FSs. 
We comment below on the case when
superconducting gaps on hole and electron FSs have different magnitudes. 
 
The free energy ${\cal F}(\Delta,M)$ to the fourth order in parameters ($\Delta$, $M$) 
 can be written in the form 
\be {\cal F}=-\alpha_s \Delta^2-\alpha_m M^2+A\Delta ^4+B
M^4+2C\Delta^2M^2. 
\label{eq:F} 
\ee 
Below we find from microscopic considerations 
how all five prefactors in \req{eq:F} depend on doping and ellipticity. 
But for the moment, we simply assume that $\alpha_s$, $A$, $B$, and $C$
 are positive constants and $\alpha_m$ varies from 
%%% that initial value of is larger than
$\alpha_m>\alpha_s$ to $\alpha_m=0$. 
This mimics the system behavior upon doping, see Section III. 

The free energy extrema are found for three different cases.
\begin{itemize}
\item Pure SC state, $M=0$, $\partial {\cal F}/\partial \Delta=0$, is given by 
\be
\label{eq:Fs}
{\cal F}_s = -\frac{\alpha_s^2}{4A}, \quad \Delta^2 = \frac{\alpha_s}{2A} .
\ee
%%% (ABV - WHAT IS Fs in terms of Delta - we refer to it later?) 
\item Pure SDW state, 
 $\Delta=0$,  $\partial {\cal F}/\partial M=0$, is given by
\be
\label{eq:Fm}
{\cal F}_m  = -\frac{\alpha_m^2}{4B},\quad M^2 = \frac{\alpha_m}{2B}.
\ee
\item Mixed SC+SDW state, 
 $\Delta\neq 0$, $M\neq 0$, found from conditions $\partial {\cal F}/\partial \Delta=\partial {\cal F}/\partial M=0$, 
which, when solved for $\Delta$ and $M$, give 
\be
 M^2=\frac{\alpha_m A-\alpha_s C}{2(AB-C^2)},\quad 
 \Delta^2= \frac{\alpha_s B-\alpha_m C}{2(AB-C^2)} \,.
\label{eq:sol}
\ee
 The corresponding free energy is
\be
\label{eq:Fmix}
\begin{split}
{\cal F}_{c} &  =  {\cal F}_s-\frac{1}{4A}\frac{(\alpha_m A-\alpha_s C)^2}{AB-C^2}\\
& = {\cal F}_m-\frac{1}{4B}\frac{(\alpha_s B-\alpha_m C)^2}{AB-C^2}.
\end{split}
\ee
\end{itemize}
These solutions are only meaningful when both $M^2>0$ and $\Delta^2>0$.

To describe 
 the system behavior with decreasing $\alpha_m$, we also introduce
 the free energy ${\cal F} (M)$
 as a function of $M$ only, obtained from \req{eq:F} 
by substituting 
%%% in the extremal value of 
\be 
\label{eq:path}
\Delta^2(M) = \frac{\alpha_s}{2A} -\frac{C}{A} M^2 \,.
\ee 
as a solution to $\partial {\cal F}/\partial \Delta=0$. 
 This gives
\be 
\label{eq:FDelta}
\begin{split}
{\cal F} (M) & =  {\cal F}_s +\gamma M^2 + \beta M^4, \\
&\gamma = -\alpha_m + \alpha_s C/A,\quad \beta = B - C^2/A.
\end{split}
\ee
The condition $\Delta^2(M) \geq 0$ determines the upper limit on a value of the SDW order, 
$M^2_{m} = \alpha_s/(2C)$.  
 The extremum of ${\cal F} (M)$, if it
exists below $M_{m}$, yields the free energy in the coexistence  state, 
${\cal F}_{c}$.  Equation~\rref{eq:Fmix} shows that the free energy in the mixed state is 
the lowest of the three 
 when $\chi \equiv AB-C^2$ is positive, or $\beta >0$
in \req{eq:FDelta}.

\begin{figure}
\centerline{\includegraphics[width=0.95\linewidth]{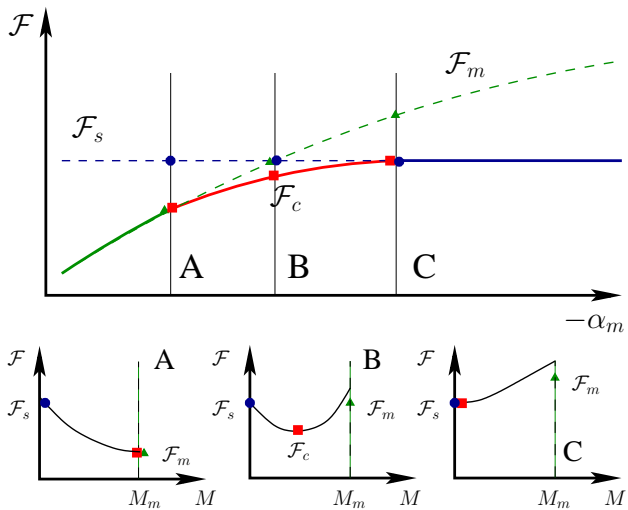}}
\caption{(Color online)
Top: 
 Evolution of the 
extreme values of the free energy, ${\cal F}_{s,c,m}$ as a function of $\alpha_m$ for $\chi=AB-C^2>0$. When the local extremum for $\Delta\neq 0$ and $M\neq 0$ appears, it becomes the global minimum and describes a thermodynamically stable phase with coexisting SC and SDW orders. Bottom: Dependence of the free energy along the trajectory  $\Delta(M)$, see Eq. (\req{eq:path}), 
and the values of ${\cal F}_s$ ($\bigcirc$), ${\cal F}_m$ ($\triangle$) and ${\cal F}_c$ ($\square$). (A) $\alpha_m=\alpha_s B/C$, when a small SC order appears within SDW phase,
  ${\cal F}_m={\cal F}_c$; 
$\alpha_m=\alpha_s\sqrt{B/A}$, when the mixed phase is the global minimum,
  ${\cal F}_m={\cal F}_s>{\cal F}_c$;
 and (C) $\alpha_m=\alpha_s C/A$, when SDW order disappears,
 ${\cal F}_s={\cal F}_c$.}
\label{fig:1}
\end{figure}

First, we consider the case $\chi>0$.
Combining $AB-C^2 >0$ with the conditions that the mixed phase only exists when
$\Delta^2 >0$ and $M^2 >0$, we show the behavior of values ${\cal F}_{s,m,c}$ 
as functions of $\alpha_m$ in top panel of Fig.~\ref{fig:1}. We
also plot ${\cal F} (M)$ for three values of  $\alpha_m$: (A)
$\alpha_m=\alpha_s B/C$, (B) $\alpha_m=\alpha_s\sqrt{B/A}$ and (C)
$\alpha_m=\alpha_s C/A$.
 The triangle symbol ($\triangle$) in these plots at $M = M_{m}$ is the value of ${\cal F}_m$ 
 from Eq. (\ref{eq:Fm}) corresponding to the extremum of the free energy
 at $\Delta\equiv 0$.
 In  general, ${\cal F}_m$  is different  from  ${\cal F} (M_m)$, both of which are obtained from \req{eq:F} with $\Delta=0$, but $M=\sqrt{\alpha_m/2B}$ or $M=\sqrt{\alpha_s/2C}$, respectively.

At large  $\alpha_m > \alpha_s B/C$, the system is in the SDW phase and 
${\cal F} (M)$  monotonically decreases with increasing $M$ to ${\cal F}(M_m)$, but ${\cal F}_m$
has even smaller value than ${\cal F}(M_m)$.  In this case $\gamma<0$ and
$\beta>0$ in Eq. (\ref{eq:FDelta}).  At such $\alpha_m$ the solution
corresponding to the coexistence state does not exist because the 
condition $\Delta^2>0$ is not yet satisfied.  At  $\alpha_m \leq \alpha_s B/C$, 
%%% $\Delta^2$ becomes positive, 
and the minimum of ${\cal F}(M)$ takes place at $M$, given by
\req{eq:sol}, for  $0<M<M_m$ when the coexistence state develops (still, we
have $\gamma<0$ and $\beta>0$ in Eq. (\ref{eq:FDelta}). The minimum splits from
$\Delta =0$ at $\alpha_m=\alpha_sB/C$ via a continuous second-order
transition. When $\alpha_m$ decreases further, but still
$\alpha_m>\alpha_sC/A$, the minimum shifts to smaller $M$, and eventually, at
$\alpha_m =\alpha_s C/A$, reaches the value ${\cal F}_s$ at $M=0$. 
 At even smaller $\alpha_m$, the global minimum corresponds to ${\cal F}_s$
 i.e., the system gradually transforms from the mixed state into the
 superconducting state.  The overall evolution of the system has 
  two second order transitions at $\alpha_m = \alpha_s B/C$ and 
 $\alpha_m = \alpha_s C/A$, and the intermediate mixed state at 
 $\alpha_s C/A < \alpha_m < \alpha_s B/C$. 

\begin{figure}
\centerline{\includegraphics[width=0.95\linewidth]{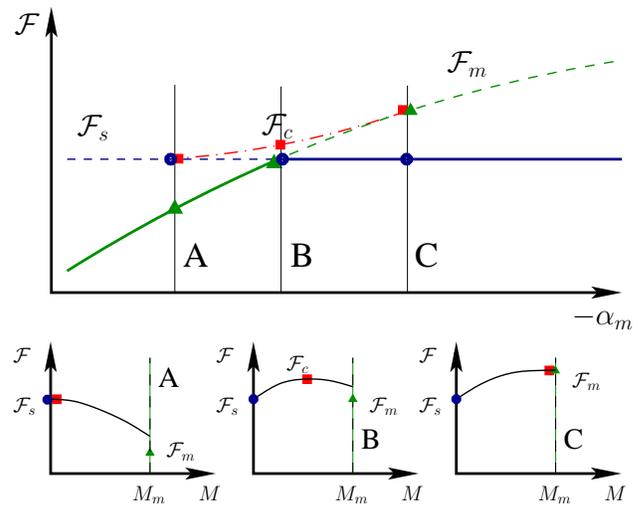}}
\caption{(Color online)
Top: Evolution of the extreme values of the free energy, ${\cal F}_{s,c,m}$ as a function of $\alpha_m$ for $\chi=AB-C^2<0$. The local extremum ${\cal F}_{c}$ 
for $\Delta\neq 0$ and $M\neq 0$  corresponds to a local  maximum and 
represents thermodynamically unstable state. 
 Bottom: Dependence of the free energy along the trajectory
 $\Delta(M)$, (see Eq. \req{eq:path}), and the values of ${\cal F}_s$ ($\bigcirc$), ${\cal F}_m$ ($\triangle$) and ${\cal F}_c$ ($\square$). (A) $\alpha_m=\alpha_s C/A$. The local extremum ${\cal F}_s$ appears at this point, 
 but the pure SDW state has a smaller energy ${\cal F}_s={\cal F}_c>{\cal F}_m$; (B)
$\alpha_m=\alpha_s\sqrt{B/A}$. At this point,  ${\cal F}_m={\cal F}_s<{\cal F}_c$, and the system undergoes a  first order transition between SC and SDW states,
(C) $\alpha_m=\alpha_s C/A$. The local extremum now corresponds to a  
weak SC order,  but the pure SC state has a lower free energy, ${\cal F}_m={\cal F}_c>{\cal F}_s$.}
\label{fig:2}
\end{figure}

In the opposite case, $AB-C^2<0$, the free energy of the ``coexistence''  state is
always larger than the free energy of pure SDW or SC
states, i.e, the ``coexistence'' state corresponds to a saddle point of the
free energy and does not represent an actual thermodynamic state of the system.
%%% The only exception is when $\alpha_s = \alpha_m B/C$, and ${\cal F}_{c}  = {\cal F}_s$. 
The evolution of ${\cal F}_{s,m,c}$ is 
  shown in Fig.~\ref{fig:2}.
 At large $\alpha_m>\alpha_sC/A$, the free energy reaches its minimum at
 $\Delta =0$, $M^2 =\alpha_m/(2B)$ and the system is in the pure SDW phase.
In this case, $\gamma >0$ and $\beta <0$ in \req{eq:FDelta}.  The
 ``coexistence'' state solution does not exist because it formally corresponds
 to $\Delta^2 <0$. At $\alpha_m = \alpha_s  C/A$,  ${\cal F}_c$ and ${\cal
 F}_s$ coincide.
  At smaller $\alpha_m$  the mixed state solution becomes real, in the sense
 that it  corresponds to $\Delta^2 >0$, and  ${\cal F} (M)$ develops a
 maximum at $M$, defined by  \req{eq:sol}. As $\alpha_m$ decreases, the maximum moves
 to larger $M$, and simultaneously ${\cal F}_m$ increases.  At
 $\alpha_m=\alpha_s\sqrt{B/A}$,  the free energies for pure states, ${\cal F}_m
 $ and  ${\cal F}_s$ become equal, while the ``coexistence'' state has a higher
 energy.  At even smaller $\alpha_m$,   ${\cal F}_m>{\cal F}_s$ and the pure SC
 phase is a true thermodynamic equilibrium state. The magnetic state remains a
 local minimum of ${\cal F}(\Delta,M)$ and the ``coexistence'' state remains a local maximum of ${\cal
 F}(M)$ down to $\alpha_m = \alpha_s B/C$, at even smaller $\alpha_m$
%AC 
coefficient $\gamma$ changes sign
 and  the ``coexistence'' solution disappears.

\section {Application to pnictides}
 
We now apply the above analysis of the two-parameter model to the pnictide
superconductors. We approximate the electronic structure of
pnictides by a  model of two  families of fermions, which form one hole and one
electron FSs of approximately equal sizes.  We assume that the hole
FS is circular, and the dispersion of fermions near this FS is 
\be
\varepsilon_h = \mu_h - \frac{k^2}{2m_h}.
\ee
The electronic FS is an ellipse, and the dispersion of fermions near this FS is \bea
&&\varepsilon_e = -\mu_e + \frac{k^2_x}{2m_x} + \frac{k^2_y}{2m_y}  \nonumber \\ 
&& = - \varepsilon_h + \mu_h - \mu_e + \frac{k^2}{2} \left[\frac{(m_x + m_y)}{2m_xm_y} -\frac{1}{m_h}\right] \nonumber\\
&& + \frac{k^2_x-k^2_y}{2} \frac{m_y-m_x}{2m_x m_y}.
\label{eq:xie}
\eea 
The last three terms represent different deviations from perfect nesting: \emph{i}) the change in chemical potentials $\delta \mu=\mu_h-\mu_e$, 
\emph{ii}) the difference in the electron and hole masses, $m_{x,y}\neq m_h$, and \emph{iii}) ellipticity, $m_x\neq m_y$. 
We will see that typical $\varepsilon_h$ are of order of temperature $T$.
 We assume that the chemical potential $\mu$ is much larger than $T$ and
neglect all terms arising due to deviations from prefect nesting with contribution to the free energy 
small in parameter $T/\mu_h$.
Within this approximation, we can set $k = k_F=\sqrt{2m_h \mu_h }$ in the two terms in the last line of Eq.   (\ref{eq:xie}). Then $\varepsilon_e = - \varepsilon_h + 2\delta_\varphi$, where
\bea
\delta_\varphi & = & \delta_0+\delta_2 \cos 2\varphi, \quad\delta_2=\frac{k_F^2}{8}\frac{m_y-m_x}{m_xm_y}\, ,\\
 \delta_0 & = & \frac{\mu_h-\mu_e}{2}+\frac{k_F^2}{4}\left(\frac{m_x+m_y}{2m_xm_y}-\frac{1}{m_h}\right).
\label{eq:delta}
\eea

Within this approximation, the two dispersions differ by a term $2\delta_{\varphi}$ which depends on the angle along the FS but does not depend on $\varepsilon_h$. One can verify that with these $\varepsilon_h$ and $\varepsilon_e$ superconducting gaps along the hole and electron FSs are equal in magnitude and different in sign, if,
  indeed, 
 the pairing interaction is approximated by a constant and $\mu$ is set to be much larger than $T$. 

The free energy for the case when the two dispersions differ by a constant was
presented in Ref.~\onlinecite{anton} for circular FSs, when $\delta_\varphi = \delta_0$
is just a constant.  Extending the expression for the free energy to the case
when $\delta_\varphi$ depends on $\varphi$ due to ellipticity, we obtain  

\be 
\label{eq:F2band}
\begin{split}
\hspace*{-1cm}
&\frac{{\cal F}(\Delta, M)}{4N_f} =
\frac{|\Delta|^2}{2} \ln\frac{T}{T_s} + \frac{ M^2}{2} \ln\frac{T}{T_m}+ 
\pi T \sum_{\vare_m} \frac{|\Delta|^2+M^2}{2|\vare_m|}
\\
&-
\pi T \sum_{\vare_m} Re\left\langle
\sqrt{(E_m+i\delta_\varphi)^2+M^2} - |\vare_m| 
\right\rangle_\varphi,
\end{split}
\ee
where $E_m=\sqrt{\vare_m^2+\Delta^2}$, $\vare_m=\pi T(2m+1)$ are the Matsubara
frequencies,  and $\langle\dots\rangle_\varphi$ imply averaging over the
direction $\varphi$ on the Fermi surface.  Temperatures $T_s$ and  $T_m$ are
transition temperatures to SC  or SDW states for ``pure'' cases when
interactions are exclusively in the SC or  SDW channels. %ABV- are present. 

Expanding the free energy,  \req{eq:F2band}, up to the fourth order 
in $M$ and $\Delta$ and comparing the result with Eq. (\ref{eq:F})  we obtain
\bea 
A& = &  \frac{ \pi T}{4}\sum_{m>0}\frac{1}{\vare_m^3},
\label{eq:A}\\
B&  =&  \frac{\pi T}{4}\sum_{m>0}\left\langle \vare_m
\frac{\vare_m^2-3 \delta_\varphi^2}{(\vare_m^2+\delta_\varphi^2)^3}\right\rangle,
\label{eq:B}
\\
C&= &\frac{\pi T}{4}\sum_{m>0}\left\langle 
\frac{\vare_m^2-  \delta_\varphi^2}{\vare_m(\vare_m^2+\delta_\varphi^2)^2}\right\rangle\,.
\label{eq:C}
\eea
and 
\bea
\alpha_s & = & \frac{1}{2}\ln(T_s/T),  \\
\alpha_m & = & \frac{1}{2}\ln\frac{T_m}{T}-2\pi T\sum_{m>0}\left\langle\frac{\delta_\varphi^2}{\vare_m(\vare_m^2+\delta_\varphi^2)}\right\rangle .
\eea
The superconducting part of the free energy, expressed via $\alpha_s$ and $A$,
is independent on $\delta_\varphi$, but the magnetic part and the mixed
$\Delta^2 M^2$ term  depend on $\delta_{\varphi}$. 
The expansion makes sense if $\Delta$, $M$ are 
of the same order, i.e. if 
$T_s$ and $T_m$ do not differ much, which we assume henceforth. 

For perfect nesting $\delta_\varphi\equiv 0$ and we have $A=B=C$, i.e., $\chi =0$, as was 
 explicitly stated in Ref.~\onlinecite{nature}.
Furthermore, the interaction term in the free energy is 
$A(\Delta^2+M^2)^2$
 and does not favor either SC or SDW orders
 even beyond the expansion to the fourth order.~\cite{anton}  The transition occurs
 into a state with a higher transition temperature $T_s$ or $T_m$, and once
 either SDW or SC order develops, the other order does not appear simply
 because the quadratic term favors either $\Delta=0$ or $M=0$, and the
 interaction term is isotropic.  If $T_s=T_m$, the free energy given by Eq.
 (\ref{eq:F2band}) is $SO(5)$ symmetric, and extra terms are needed to break
 this symmetry. 

As we said, we consider the case when $T_m>T_s$, such that at perfect nesting
the system develops an SDW order.  Deviations from perfect nesting lead to two
effects. First, the magnitude of  $\alpha_m$ is reduced  which is the
manifestation of the fact that SDW instability is  suppressed
 when nesting becomes non-perfect.
  Superconducting $\alpha_s$ is not affected by  $\delta_\varphi$, and at large
  $\delta_\varphi$
 superconductivity is the only possible state. 
 Second, coefficients $B$ and $C$ evolve with
 $\delta_\varphi$, and $\chi =AB-C^2$ becomes non-zero when $\delta_\varphi$ is
 finite.  The question is what is the sign of this term. We recall that, when
 $\chi >0$, the system evolves from SDW to SC via two second order transitions
 and the intermediate coexistence phase, while for $\chi <0$, there is no mixed
 state and SDW and SC phases are separated by a first-order transition.  

To get an insight on how $\chi$ evolves at nonzero $\delta_\varphi$, we
 first consider $\delta_\varphi$ as a small parameter and 
expand $A,~B$, and $C$  in powers of $\delta_\varphi$. Collecting terms up to fourth order in $\delta_\varphi$, we obtain 
\be
\chi=\frac{1}{32\pi^8T^8}\left(
s_1\langle\delta_\varphi^4\rangle-s_2 \langle\delta_\varphi^2\rangle^2
\right),
\ee
where
\bea
&&s_1 = 5 \left(\sum_{m>0} \frac{1}{(2m+1)^3} \right)\left(  \sum_{m>0} \frac{1}{(2m+1)^7} \right),
\nonumber\\
&&s_2 = 9 \left(\sum_{m>0} \frac{1}{(2m+1)^5}\right)^2.  
\eea
The sums are expressed in terms of the Riemann-Zeta function and 
 give  $s_1\approx 5.26$ and $s_2\approx 9.08$.
Substituting $\delta_\varphi$ from (\ref{eq:delta}) and averaging, we obtain
\be
\chi\approx \frac{1}{32\pi^8T^8}\left(
-3.82\delta^4_0 +6.70 \delta^2_0 \delta^2_2 - 0.30 \delta^4_2\right).
\label{eq:chi_exp}
\ee
We emphasize that in the two limits when
either $\delta_2 =0$ or $\delta_0=0$, $\chi <0$, i.e., the transition is first
order. The first limit corresponds to circular FSs with different chemical
potentials, the second limit corresponds to the case when chemical potential remain
equal but electron dispersion becomes elliptical.  
In both
cases, SDW order opens gaps for some fermionic excitations, but still
preserves low-energy fermionic states near the modified FSs. Fermions near
these FSs  still have a non-zero $s^{+-}$ SC solution, however, this solution represents an
energetically unfavorable state. 
 We particularly emphasize that the ellipticity of
electron dispersion, taken alone (i.e., $\delta_0=0$) is not sufficient for appearance 
of the coexistence phase. 
 
\begin{figure}
\centerline{\includegraphics[width=0.8\linewidth]{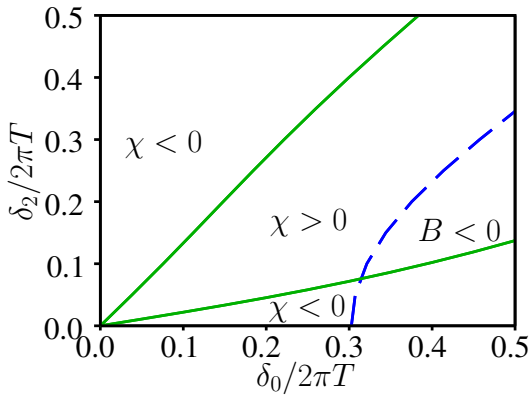}}
\caption{(Color online)
Contour plot of the sign of $\chi = AB-C^2$, where $A$, $B$, and $C$ are given by  Eqs. (\protect\ref{eq:A})-(\protect\ref{eq:C}). The area between the two solid lines 
corresponds to positive values of $\chi$, where the coexistence 
between the SDW and SC phases is possible. Below the dashed line the 
value of $B$ is negative, and the system 
 undergoes a 
transition from the normal state to the incommensurate SDW state.
The mixed phase between SC and incommensurate SDW orders does exist,\protect\cite{anton} but its boundaries are shifted compared to the ones for commensurate SDW.}
\label{fig:phase}
\end{figure}

When both $\delta_0$ {\it and} $\delta_2$ are non-zero, there is a
relatively broad range $0.76<\delta_2/\delta_0<4.68$  where $\chi >0$ and the
transformation from pure SDW to SC phases occurs via a coexistence phase.  To verify that
this statement holds at larger values of $\delta_0$ and $\delta_2$, we computed 
 $\chi = \chi(\delta_0,\delta_2)$ without expanding in $\delta_\varphi$. We
 plot the sign of $\chi (\delta_0,\delta_2)$ in Fig. \ref{fig:phase}.  We
 obtained the same result as above, namely,  for $\delta_0 =0$ or $\delta_2
 =0$, $\chi <0$ and the transition between SDW and SC states is of first order,
 while when both $\delta_0$ and $\delta_2$ are non-zero, there exists a region
 with $\chi(\delta_0,\delta_2) >0$.
 In this parameter range, the transformation from SDW to SC phases
 involves a coexistence phase.

Two remarks are in order here. In our analysis of the free energy we assumed
that 
$B$ given by \req{eq:B} is positive.  This is, however, only true if
$\delta_0$ and $\delta_2$ are below certain thresholds, see the dashed line in Fig.~\ref{fig:phase}. At larger $\delta_0$
and $\delta_2$, the coefficient $B$ becomes negative and the analysis has to be modified. 
This is the condition where  instability 
with respect to formation of an incommensurate SDW order occurs, much like 
Fulde-Ferrel-Larkin-Ovchinikov phase in superconductors.
In particular, for  $\delta_2=0$, 
 $B$ becomes negative for $\delta_0/2\pi T \approx 0.304$.~\cite{Cvetkovic09,anton}  This order develops at a temperature  above $T_s$ if 
$T_s < 0.56\,T_m$.  
We also note that Eq. (\ref{eq:chi_exp}) is obtained
under the assumption 
 that $T\ll \mu$, which allowed us to restrict 
the contributions to  $T^4 \chi$ from terms that scale as
$(\delta_\varphi/T)^4$ and neglect terms which scale as powers of
$(\delta_\varphi/\mu_h)^2$. It is unlikely but, in principle, possible that the
expansion of the full $\chi$ in powers of $\delta_\varphi$ begins with the
quadratic term $(1/T^4) (\delta_\varphi/\mu_h)^2$, i.e., $T^4 \chi = c_2
(\delta_\varphi/\mu_h)^2 + c_4 (\delta_\varphi/T)^4 +...$. If this is the case
and $c_2>0$, the mixed phase exists  in a tiny range of $\delta_\varphi$, where $\chi <0 $ without the $c_2$ term.

\section{Conventional two-band $s-$superconductivity}

For completeness, we also consider the case when superconducting order
parameter in a conventional $s-$wave, i.e., the sign of $\Delta$ is the same on
hole and electron FSs. This case has been considered in Ref.~\onlinecite{nature}, and
the conclusion was that the transition from SDW to SC phase is always of first order,
regardless of ellipticity or shift of chemical potentials.  This result
implies that the very coexistence between SDW and SC in Fe-pnictides is an
implication that the pairing state is not a conventional $s^{++}$ state.   

The analysis in Ref.~\onlinecite{nature} was based on the observation that for
$s^{++}$ gap, $\chi <0$ already for circular FSs and perfect nesting, and on
numerical calculations of $\chi$ for some cases when nesting is not perfect.
We analyze this issue analytically. 

The free energy for $s^{++}$ SC gap can be derived using the 
same approach as in Ref.~\onlinecite{anton} and has the form:
\begin{widetext}
\be 
\label{eq:F2band_2}
\begin{split}
\hspace*{-1cm}
&\frac{{\cal F}^{++}(\Delta, M)}{4N_f} =
\frac{|\Delta|^2}{2} \ln\frac{T}{T_s} + \frac{ M^2}{2} \ln\frac{T}{T_m}+ 
\pi T \sum_{\vare_m} \frac{|\Delta|^2+M^2}{2|\vare_m|}
\\
&-
\frac{\pi T}{2} \sum_{\vare_m} \sum_\pm\left\langle
\sqrt{(\vare_m^2+\Delta^2+M^2 -\delta_\varphi^2\pm 2\sqrt{\Delta^2M^2-\delta_\varphi^2(\vare_m^2+\Delta^2)}} - |\vare_m| 
\right\rangle_\varphi.
\end{split}
\ee
\end{widetext}
Expanding this free energy in powers of $\Delta^2$ and $M^2$ and comparing the
result with Eq.~(\ref{eq:F}),  we find that coefficients $A$ and $B$ are still
given by \reqs{eq:A} and \rref{eq:B}, but the coefficient $C$ is
 modified to

\be
\label{eq:Cpp}
C = \frac{\pi T}{4}\sum_{m>0}\left\langle 
\frac{3 \vare_m^2+  \delta_\varphi^2}{\vare_m(\vare_m^2+\delta_\varphi^2)^2}\right\rangle\,.
\ee
We now have 
$
\chi=Y(\delta_0,\delta_2)/\pi^4T^4 
$.
At perfect nesting, $Y(0,0)=-(7\zeta(3)/8)^2/2<0$ and the
transition  is of first order.  When $\delta_{0,2}$ increase, the magnitude of
$\chi\propto Y(\delta_0,\delta_2)$ is reduced, but we found that its sign remains negative for arbitrary
values of $\delta_0$ and $\delta_2$.  In Fig. 4, we show the behavior of $Y(\delta_0,\delta_2)$
for three cases: $Y(\delta,0)$, $Y(0,\delta)$, and $Y(\delta,\delta)$. We see that
in all cases $Y(\delta_0,\delta_2)$ remains negative for arbitrary magnitudes of 
$\delta_{0,2}$. We computed the sign of $\chi$ everywhere in
the $(\delta_0,\delta_2)$ plane and found that it is always negative. Therefore, we 
confirm the result of Ref.~\onlinecite{nature} that there is no mixed phase if the SC
gap has $s^{++}$ symmetry on both FSs.

\begin{figure}
\centerline{\includegraphics[width=0.95\linewidth]{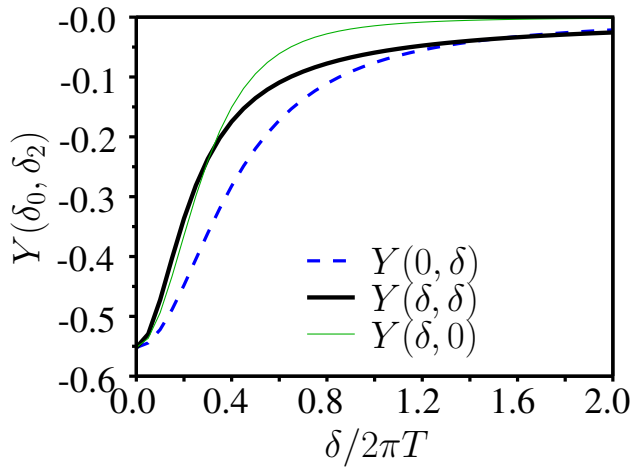}}
\caption{(Color online)  The behavior of $Y(\delta_0,\delta_2)$
for three cases: $Y(\delta,0)$, $Y(0,\delta)$, and $Y(\delta,\delta)$. 
In all cases $Y(\delta_0,\delta_2)$ monotonically increases with increasing argument, but remains negative.}
\label{fig:4}
\end{figure}

\begin{figure}
\centerline{\includegraphics[width=0.95\linewidth]{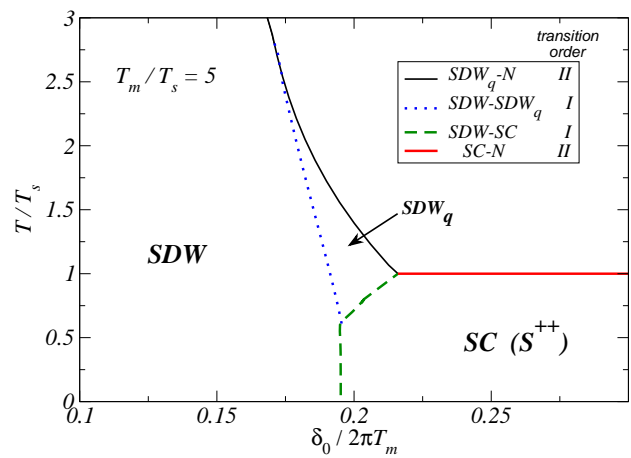}}
\caption{(Color online)
The phase diagram for $T_m/T_s=5$ for  $s^{++}$ 
 superconductivity
for two circular Fermi surfaces as a function of $\delta_0$.
 The solid curved 
 line is the transition between the normal and (in)commensurate SDW($_q$) states.  The dotted line 
  is the first order transition line  between commensurate and
incommensurate SDW phases. The solid horizontal line is the second order normal-SC transition.
  The transitions between both, incommensurate and commensurate SDW phases, 
and SC phase are first order (dashed lines), and the mixed phase does not appear. 
 For an $s^{+-}$ SC, the transition between mixed SC+SDW$_q$ and pure SC state is second order.~\protect\cite{anton}}
\label{fig:inc_s++}
\end{figure}

We went a step further and analyzed what happens when $T_m$ is much greater
than $T_s$, and magnetic order becomes incommensurate at high temperatures $T>T_s$. 
The results are presented in Fig.~\ref{fig:inc_s++}. We observe that, even when
 magnetic order is incommensurate, the transition between SDW and SC phases remains of first order, shown by the dashed line in Fig.~\ref{fig:inc_s++},
and the mixed state does not occur.
To the contrary, for an $s^{+-}$ SC, once  SDW order becomes incommensurate, the
transformation from incommensurate SDW to SC involves the coexistence phase even though the transition from
commensurate SDW to SC phase would be of first order.  

\section{Conclusions}

In this paper we studied analytically the interplay between a commensurate SDW
magnetism and  superconductivity  in the model for Fe-pnictides. The 
conventional wisdom is that the transition from SDW to SC should be of first order
in the case of a perfect nesting, because 
then the SDW order fully gaps electronic states leaving no space for SC, but should
involve a coexistence  phase for non-perfect nesting because then SDW order leaves
modified FSs on which a SC order may develop.
 We show that the situation is more complex, and the presence of FSs in the
 SDW phase is not the sufficient 
 condition  for the mixed state to emerge. We
 show explicitly that the transition remains of first order when doping  shifts  hole and electron bands, but the bands remain
 circular, {\it and} when the electron band becomes elliptical, but chemical
 potential does not shift (the material remains a compensated metal). Only when both ellipticity and a shift of the
 chemical potential are present, we found the mixed state in some range of
 parameters. 

We also analyzed the case when the SC order has a conventional, sign-preserving
$s^{++}$ symmetry on the two FSs and confirmed analytically the result of Ref.\onlinecite{nature}
that the transition between SDW and SC phases is always of first order, and the
mixed phase does not appear. We argue that this is the case even when the SDW order
becomes incommensurate.   

\section{Acknowledgments}

We thank  I. Eremin, R. Fernandes,
 I. Mazin,   J. Schmalian,  O. Sushkov, and  Z. Tesanovic
 for useful discussions.
 M.G.V. acknowledges the Donors of the American Chemical Society Petroleum Research Fund for partial support and A.V.C. acknowledges the support from NSF-DMR 0906953.


\begin{thebibliography}{36}
\expandafter\ifx\csname natexlab\endcsname\relax\def\natexlab#1{#1}\fi
\expandafter\ifx\csname bibnamefont\endcsname\relax
  \def\bibnamefont#1{#1}\fi
\expandafter\ifx\csname bibfnamefont\endcsname\relax
  \def\bibfnamefont#1{#1}\fi
\expandafter\ifx\csname citenamefont\endcsname\relax
  \def\citenamefont#1{#1}\fi
\expandafter\ifx\csname url\endcsname\relax
  \def\url#1{\texttt{#1}}\fi
\expandafter\ifx\csname urlprefix\endcsname\relax\def\urlprefix{URL }\fi
\providecommand{\bibinfo}[2]{#2}
\providecommand{\eprint}[2][]{\url{#2}}

\bibitem[{\citenamefont{Kamihara et~al.}({2008})\citenamefont{Kamihara,
  Watanabe, Hirano, and Hosono}}]{Kamihara08}
\bibinfo{author}{\bibfnamefont{Y.}~\bibnamefont{Kamihara}},
  \bibinfo{author}{\bibfnamefont{T.}~\bibnamefont{Watanabe}},
  \bibinfo{author}{\bibfnamefont{M.}~\bibnamefont{Hirano}}, \bibnamefont{and}
  \bibinfo{author}{\bibfnamefont{H.}~\bibnamefont{Hosono}},
  \bibinfo{journal}{{J. Am. Chem. Soc.}} \textbf{\bibinfo{volume}{{130}}},
  \bibinfo{pages}{3296} (\bibinfo{year}{{2008}}).

\bibitem[{\citenamefont{Luetkens et~al.}({2009})\citenamefont{Luetkens, Klauss,
  Kraken, Litterst, Dellmann, Klingeler, Hess, Khasanov, Amato, Baines
  et~al.}}]{Luetkens09}
\bibinfo{author}{\bibfnamefont{H.}~\bibnamefont{Luetkens}},
  \bibinfo{author}{\bibfnamefont{H.-H.} \bibnamefont{Klauss}},
  \bibinfo{author}{\bibfnamefont{M.}~\bibnamefont{Kraken}},
  \bibinfo{author}{\bibfnamefont{F.~J.} \bibnamefont{Litterst}},
  \bibinfo{author}{\bibfnamefont{T.}~\bibnamefont{Dellmann}},
  \bibinfo{author}{\bibfnamefont{R.}~\bibnamefont{Klingeler}},
  \bibinfo{author}{\bibfnamefont{C.}~\bibnamefont{Hess}},
  \bibinfo{author}{\bibfnamefont{R.}~\bibnamefont{Khasanov}},
  \bibinfo{author}{\bibfnamefont{A.}~\bibnamefont{Amato}},
  \bibinfo{author}{\bibfnamefont{C.}~\bibnamefont{Baines}},
  \bibnamefont{et~al.}, \bibinfo{journal}{{Nature Materials}}
  \textbf{\bibinfo{volume}{{8}}}, \bibinfo{pages}{305}
  (\bibinfo{year}{{2009}}).

\bibitem[{\citenamefont{Laplace et~al.}({2009})\citenamefont{Laplace, Bobroff,
  Rullier-Albenque, Colson, and Forget}}]{Laplace09}
\bibinfo{author}{\bibfnamefont{Y.}~\bibnamefont{Laplace}},
  \bibinfo{author}{\bibfnamefont{J.}~\bibnamefont{Bobroff}},
  \bibinfo{author}{\bibfnamefont{F.}~\bibnamefont{Rullier-Albenque}},
  \bibinfo{author}{\bibfnamefont{D.}~\bibnamefont{Colson}}, \bibnamefont{and}
  \bibinfo{author}{\bibfnamefont{A.}~\bibnamefont{Forget}},
  \bibinfo{journal}{Phys. Rev. B} \textbf{\bibinfo{volume}{{80}}},
  \bibinfo{pages}{140501} (\bibinfo{year}{{2009}}).

\bibitem[{\citenamefont{Chu et~al.}({2009})\citenamefont{Chu, Analytis,
  Kucharczyk, and Fisher}}]{Chu09}
\bibinfo{author}{\bibfnamefont{J.-H.} \bibnamefont{Chu}},
  \bibinfo{author}{\bibfnamefont{J.~G.} \bibnamefont{Analytis}},
  \bibinfo{author}{\bibfnamefont{C.}~\bibnamefont{Kucharczyk}},
  \bibnamefont{and} \bibinfo{author}{\bibfnamefont{I.~R.}
  \bibnamefont{Fisher}}, \bibinfo{journal}{Phys. Rev. B}
  \textbf{\bibinfo{volume}{{79}}}, \bibinfo{pages}{014506}
  (\bibinfo{year}{{2009}}).

\bibitem[{\citenamefont{Rotter et~al.}({2008})\citenamefont{Rotter, Tegel, and
  Johrendt}}]{Rotter08}
\bibinfo{author}{\bibfnamefont{M.}~\bibnamefont{Rotter}},
  \bibinfo{author}{\bibfnamefont{M.}~\bibnamefont{Tegel}}, \bibnamefont{and}
  \bibinfo{author}{\bibfnamefont{D.}~\bibnamefont{Johrendt}},
  \bibinfo{journal}{Phys. Rev. Lett.} \textbf{\bibinfo{volume}{{101}}},
  \bibinfo{pages}{107006} (\bibinfo{year}{{2008}}).

\bibitem[{\citenamefont{Fernandes et~al.}({2009})\citenamefont{Fernandes,
  Pratt, Tian, Zarestky, Kreyssig, Nandi, Kim, Thaler, Ni, Bud’ko
  et~al.}}]{nature}
\bibinfo{author}{\bibfnamefont{R.~M.} \bibnamefont{Fernandes}},
  \bibinfo{author}{\bibfnamefont{D.~K.} \bibnamefont{Pratt}},
  \bibinfo{author}{\bibfnamefont{W.}~\bibnamefont{Tian}},
  \bibinfo{author}{\bibfnamefont{J.}~\bibnamefont{Zarestky}},
  \bibinfo{author}{\bibfnamefont{A.}~\bibnamefont{Kreyssig}},
  \bibinfo{author}{\bibfnamefont{S.}~\bibnamefont{Nandi}},
  \bibinfo{author}{\bibfnamefont{M.~G.} \bibnamefont{Kim}},
  \bibinfo{author}{\bibfnamefont{A.}~\bibnamefont{Thaler}},
  \bibinfo{author}{\bibfnamefont{N.}~\bibnamefont{Ni}},
  \bibinfo{author}{\bibfnamefont{S.~L.} \bibnamefont{Bud’ko}},
  \bibnamefont{et~al.}, \bibinfo{howpublished}{arXiv:0911.5183}
  (\bibinfo{year}{{2009}}).

\bibitem[{\citenamefont{Liu et~al.}({2008})\citenamefont{Liu, Samolyuk, Lee,
  Ni, Kondo, Santander-Syro, Bud’ko, McChesney, Rotenberg, Valla
  et~al.}}]{ARPES-Liu08}
\bibinfo{author}{\bibfnamefont{C.}~\bibnamefont{Liu}},
  \bibinfo{author}{\bibfnamefont{G.~D.} \bibnamefont{Samolyuk}},
  \bibinfo{author}{\bibfnamefont{Y.}~\bibnamefont{Lee}},
  \bibinfo{author}{\bibfnamefont{N.}~\bibnamefont{Ni}},
  \bibinfo{author}{\bibfnamefont{T.}~\bibnamefont{Kondo}},
  \bibinfo{author}{\bibfnamefont{A.~F.} \bibnamefont{Santander-Syro}},
  \bibinfo{author}{\bibfnamefont{S.~L.} \bibnamefont{Bud’ko}},
  \bibinfo{author}{\bibfnamefont{J.~L.} \bibnamefont{McChesney}},
  \bibinfo{author}{\bibfnamefont{E.}~\bibnamefont{Rotenberg}},
  \bibinfo{author}{\bibfnamefont{T.}~\bibnamefont{Valla}},
  \bibnamefont{et~al.}, \bibinfo{journal}{Phys. Rev. Lett.}
  \textbf{\bibinfo{volume}{{101}}}, \bibinfo{pages}{177005}
  (\bibinfo{year}{{2008}}).

\bibitem[{\citenamefont{Evtushinsky et~al.}({2009})\citenamefont{Evtushinsky,
  Inosov, Zabolotnyy, Koitzsch, Knupfer, Büchner, Viazovska, Sun, Hinkov,
  Boris et~al.}}]{ARPES-Evtushinsky09}
\bibinfo{author}{\bibfnamefont{D.~V.} \bibnamefont{Evtushinsky}},
  \bibinfo{author}{\bibfnamefont{D.~S.} \bibnamefont{Inosov}},
  \bibinfo{author}{\bibfnamefont{V.~B.} \bibnamefont{Zabolotnyy}},
  \bibinfo{author}{\bibfnamefont{A.}~\bibnamefont{Koitzsch}},
  \bibinfo{author}{\bibfnamefont{M.}~\bibnamefont{Knupfer}},
  \bibinfo{author}{\bibfnamefont{B.}~\bibnamefont{Büchner}},
  \bibinfo{author}{\bibfnamefont{M.~S.} \bibnamefont{Viazovska}},
  \bibinfo{author}{\bibfnamefont{G.~L.} \bibnamefont{Sun}},
  \bibinfo{author}{\bibfnamefont{V.}~\bibnamefont{Hinkov}},
  \bibinfo{author}{\bibfnamefont{A.~V.} \bibnamefont{Boris}},
  \bibnamefont{et~al.}, \bibinfo{journal}{Phys. Rev. B}
  \textbf{\bibinfo{volume}{{79}}}, \bibinfo{pages}{054517}
  (\bibinfo{year}{{2009}}).

\bibitem[{\citenamefont{Hsieh et~al.}({2008})\citenamefont{Hsieh, Xia, Wray,
  Qian, Gomes, Yazdani, Chen, Luo, Wang, and Hasan}}]{ARPES-Hsieh08}
\bibinfo{author}{\bibfnamefont{D.}~\bibnamefont{Hsieh}},
  \bibinfo{author}{\bibfnamefont{Y.}~\bibnamefont{Xia}},
  \bibinfo{author}{\bibfnamefont{L.}~\bibnamefont{Wray}},
  \bibinfo{author}{\bibfnamefont{D.}~\bibnamefont{Qian}},
  \bibinfo{author}{\bibfnamefont{K.}~\bibnamefont{Gomes}},
  \bibinfo{author}{\bibfnamefont{A.}~\bibnamefont{Yazdani}},
  \bibinfo{author}{\bibfnamefont{G.~F.} \bibnamefont{Chen}},
  \bibinfo{author}{\bibfnamefont{J.~L.} \bibnamefont{Luo}},
  \bibinfo{author}{\bibfnamefont{N.~L.} \bibnamefont{Wang}}, \bibnamefont{and}
  \bibinfo{author}{\bibfnamefont{M.~Z.} \bibnamefont{Hasan}},
  \bibinfo{howpublished}{{ArXiv:0812.2289v1}} (\bibinfo{year}{{2008}}).

\bibitem[{\citenamefont{Ding et~al.}({2008})\citenamefont{Ding, Nakayama,
  Richard, Souma, Sato, Takahashi, Neupane, Xu, Pan, Federov
  et~al.}}]{Ding08-1}
\bibinfo{author}{\bibfnamefont{H.}~\bibnamefont{Ding}},
  \bibinfo{author}{\bibfnamefont{K.}~\bibnamefont{Nakayama}},
  \bibinfo{author}{\bibfnamefont{P.}~\bibnamefont{Richard}},
  \bibinfo{author}{\bibfnamefont{S.}~\bibnamefont{Souma}},
  \bibinfo{author}{\bibfnamefont{T.}~\bibnamefont{Sato}},
  \bibinfo{author}{\bibfnamefont{T.}~\bibnamefont{Takahashi}},
  \bibinfo{author}{\bibfnamefont{M.}~\bibnamefont{Neupane}},
  \bibinfo{author}{\bibfnamefont{Y.~M.} \bibnamefont{Xu}},
  \bibinfo{author}{\bibfnamefont{Z.~H.} \bibnamefont{Pan}},
  \bibinfo{author}{\bibfnamefont{A.~V.} \bibnamefont{Federov}},
  \bibnamefont{et~al.}, \bibinfo{howpublished}{arXiv:0812.0534}
  (\bibinfo{year}{{2008}}).

\bibitem[{\citenamefont{Zabolotnyy et~al.}({2009})\citenamefont{Zabolotnyy,
  Inosov, Evtushinsky, Koitzsch, Kordyuk, Sun, Park, Haug, Hinkov, Boris
  et~al.}}]{Zabolotnyy09}
\bibinfo{author}{\bibfnamefont{V.~B.} \bibnamefont{Zabolotnyy}},
  \bibinfo{author}{\bibfnamefont{D.~S.} \bibnamefont{Inosov}},
  \bibinfo{author}{\bibfnamefont{D.~V.} \bibnamefont{Evtushinsky}},
  \bibinfo{author}{\bibfnamefont{A.}~\bibnamefont{Koitzsch}},
  \bibinfo{author}{\bibfnamefont{A.~A.} \bibnamefont{Kordyuk}},
  \bibinfo{author}{\bibfnamefont{G.~L.} \bibnamefont{Sun}},
  \bibinfo{author}{\bibfnamefont{J.~T.} \bibnamefont{Park}},
  \bibinfo{author}{\bibfnamefont{D.}~\bibnamefont{Haug}},
  \bibinfo{author}{\bibfnamefont{V.}~\bibnamefont{Hinkov}},
  \bibinfo{author}{\bibfnamefont{A.~V.} \bibnamefont{Boris}},
  \bibnamefont{et~al.}, \bibinfo{journal}{{Nature}}
  \textbf{\bibinfo{volume}{{457}}}, \bibinfo{pages}{569}
  (\bibinfo{year}{{2009}}).

\bibitem[{\citenamefont{Coldea et~al.}({2008})\citenamefont{Coldea, Fletcher,
  Carrington, Analytis, Bangura, Chu, Erickson, Fisher, Hussey, and
  McDonald}}]{Coldea08}
\bibinfo{author}{\bibfnamefont{A.~I.} \bibnamefont{Coldea}},
  \bibinfo{author}{\bibfnamefont{J.~D.} \bibnamefont{Fletcher}},
  \bibinfo{author}{\bibfnamefont{A.}~\bibnamefont{Carrington}},
  \bibinfo{author}{\bibfnamefont{J.~G.} \bibnamefont{Analytis}},
  \bibinfo{author}{\bibfnamefont{A.~F.} \bibnamefont{Bangura}},
  \bibinfo{author}{\bibfnamefont{J.~H.} \bibnamefont{Chu}},
  \bibinfo{author}{\bibfnamefont{A.~S.} \bibnamefont{Erickson}},
  \bibinfo{author}{\bibfnamefont{I.~R.} \bibnamefont{Fisher}},
  \bibinfo{author}{\bibfnamefont{N.~E.} \bibnamefont{Hussey}},
  \bibnamefont{and} \bibinfo{author}{\bibfnamefont{R.~D.}
  \bibnamefont{McDonald}}, \bibinfo{journal}{{Phys. Rev. Lett.}}
  \textbf{\bibinfo{volume}{{101}}}, \bibinfo{pages}{216402}
  (\bibinfo{year}{{2008}}).

\bibitem[{\citenamefont{Coldea et~al.}({2009})\citenamefont{Coldea, Andrew,
  Analytis, McDonald, Bangura, Chu, Fisher, and Carrington}}]{Coldea09}
\bibinfo{author}{\bibfnamefont{A.~I.} \bibnamefont{Coldea}},
  \bibinfo{author}{\bibfnamefont{C.~M.~J.} \bibnamefont{Andrew}},
  \bibinfo{author}{\bibfnamefont{J.~G.} \bibnamefont{Analytis}},
  \bibinfo{author}{\bibfnamefont{R.~D.} \bibnamefont{McDonald}},
  \bibinfo{author}{\bibfnamefont{A.~F.} \bibnamefont{Bangura}},
  \bibinfo{author}{\bibfnamefont{J.-H.} \bibnamefont{Chu}},
  \bibinfo{author}{\bibfnamefont{I.~R.} \bibnamefont{Fisher}},
  \bibnamefont{and}
  \bibinfo{author}{\bibfnamefont{A.}~\bibnamefont{Carrington}},
  \bibinfo{journal}{Phys. Rev. Lett.} \textbf{\bibinfo{volume}{{103}}},
  \bibinfo{pages}{026404} (\bibinfo{year}{{2009}}).

\bibitem[{\citenamefont{Singh and Du}({2008})}]{SinghDu-PRL100}
\bibinfo{author}{\bibfnamefont{D.~J.} \bibnamefont{Singh}} \bibnamefont{and}
  \bibinfo{author}{\bibfnamefont{M.~H.} \bibnamefont{Du}},
  \bibinfo{journal}{{Phys. Rev. Lett.}} \textbf{\bibinfo{volume}{{100}}},
  \bibinfo{eid}{{237003}} 
  (\bibinfo{year}{{2008}}).

\bibitem[{\citenamefont{Boeri et~al.}({2008})\citenamefont{Boeri, Dolgov, and
  Golubov}}]{phonons}
\bibinfo{author}{\bibfnamefont{L.}~\bibnamefont{Boeri}},
  \bibinfo{author}{\bibfnamefont{O.~V.} \bibnamefont{Dolgov}},
  \bibnamefont{and} \bibinfo{author}{\bibfnamefont{A.~A.}
  \bibnamefont{Golubov}}, \bibinfo{journal}{{Phys. Rev. Lett.}}
  \textbf{\bibinfo{volume}{{101}}}, \bibinfo{eid}{{026403}}
  (\bibinfo{year}{{2008}}).

\bibitem[{\citenamefont{Mazin et~al.}({2008})\citenamefont{Mazin, Singh,
  Johannes, and Du}}]{mazin:057003}
\bibinfo{author}{\bibfnamefont{I.~I.} \bibnamefont{Mazin}},
  \bibinfo{author}{\bibfnamefont{D.~J.} \bibnamefont{Singh}},
  \bibinfo{author}{\bibfnamefont{M.~D.} \bibnamefont{Johannes}},
  \bibnamefont{and} \bibinfo{author}{\bibfnamefont{M.~H.} \bibnamefont{Du}},
  \bibinfo{journal}{Phys. Rev. Lett.} \textbf{\bibinfo{volume}{{101}}},
  \bibinfo{eid}{{057003}} 
  (\bibinfo{year}{{2008}}).

\bibitem[{\citenamefont{Chubukov et~al.}({2008})\citenamefont{Chubukov,
  Efremov, and Eremin}}]{chubukov08}
\bibinfo{author}{\bibfnamefont{A.~V.} \bibnamefont{Chubukov}},
  \bibinfo{author}{\bibfnamefont{D.}~\bibnamefont{Efremov}}, \bibnamefont{and}
  \bibinfo{author}{\bibfnamefont{I.}~\bibnamefont{Eremin}},
  \bibinfo{journal}{{Phys. Rev. B}} \textbf{\bibinfo{volume}{{78}}},
  \bibinfo{pages}{134512} (\bibinfo{year}{{2008}}).

\bibitem[{\citenamefont{Chubukov}({2009})}]{chubukov09}
\bibinfo{author}{\bibfnamefont{A.~V.}~\bibnamefont{Chubukov}},
  \bibinfo{journal}{{Physica C: Superconductivity}}
  \textbf{\bibinfo{volume}{{469}}}, \bibinfo{pages}{640}
  (\bibinfo{year}{{2009}}).

\bibitem[{\citenamefont{Kuroki et~al.}({2008})\citenamefont{Kuroki, Onari,
  Arita, Usui, Tanaka, Kontani, and Aoki}}]{Kuroki08}
\bibinfo{author}{\bibfnamefont{K.}~\bibnamefont{Kuroki}},
  \bibinfo{author}{\bibfnamefont{S.}~\bibnamefont{Onari}},
  \bibinfo{author}{\bibfnamefont{R.}~\bibnamefont{Arita}},
  \bibinfo{author}{\bibfnamefont{H.}~\bibnamefont{Usui}},
  \bibinfo{author}{\bibfnamefont{Y.}~\bibnamefont{Tanaka}},
  \bibinfo{author}{\bibfnamefont{H.}~\bibnamefont{Kontani}}, \bibnamefont{and}
  \bibinfo{author}{\bibfnamefont{H.}~\bibnamefont{Aoki}},
  \bibinfo{journal}{Phys. Rev. Lett.} \textbf{\bibinfo{volume}{{101}}},
  \bibinfo{pages}{087004} (\bibinfo{year}{{2008}}).

\bibitem[{\citenamefont{Barzykin and Gorkov}({2008})}]{Barzykin08}
\bibinfo{author}{\bibfnamefont{V.}~\bibnamefont{Barzykin}} \bibnamefont{and}
  \bibinfo{author}{\bibfnamefont{L.~P.} \bibnamefont{Gorkov}},
  \bibinfo{journal}{{JETP Letters}} \textbf{\bibinfo{volume}{{88}}},
  \bibinfo{pages}{131} (\bibinfo{year}{{2008}}).

\bibitem[{\citenamefont{Maier et~al.}({2009})\citenamefont{Maier, Graser,
  Scalapino, and Hirschfeld}}]{Maier-PRB79}
\bibinfo{author}{\bibfnamefont{T.~A.} \bibnamefont{Maier}},
  \bibinfo{author}{\bibfnamefont{S.}~\bibnamefont{Graser}},
  \bibinfo{author}{\bibfnamefont{D.~J.} \bibnamefont{Scalapino}},
  \bibnamefont{and} \bibinfo{author}{\bibfnamefont{P.~J.}
  \bibnamefont{Hirschfeld}}, \bibinfo{journal}{Phys. Rev. B}
  \textbf{\bibinfo{volume}{{79}}}, \bibinfo{pages}{224510}
  (\bibinfo{year}{{2009}}).

\bibitem[{\citenamefont{Chubukov et~al.}({2009})\citenamefont{Chubukov,
  Vavilov, and Vorontsov}}]{CVV-PRB09}
\bibinfo{author}{\bibfnamefont{A.~V.} \bibnamefont{Chubukov}},
  \bibinfo{author}{\bibfnamefont{M.~G.} \bibnamefont{Vavilov}},
  \bibnamefont{and} \bibinfo{author}{\bibfnamefont{A.~B.}
  \bibnamefont{Vorontsov}}, \bibinfo{journal}{Phys. Rev. B}
  \textbf{\bibinfo{volume}{{80}}}, \bibinfo{eid}{{140515}}
 (\bibinfo{year}{{2009}}).

\bibitem[{\citenamefont{Thomale et~al.}({2009})\citenamefont{Thomale, Platt,
  Hu, Honerkamp, and Bernevig}}]{Thomale09}
\bibinfo{author}{\bibfnamefont{R.}~\bibnamefont{Thomale}},
  \bibinfo{author}{\bibfnamefont{C.}~\bibnamefont{Platt}},
  \bibinfo{author}{\bibfnamefont{J.}~\bibnamefont{Hu}},
  \bibinfo{author}{\bibfnamefont{C.}~\bibnamefont{Honerkamp}},
  \bibnamefont{and} \bibinfo{author}{\bibfnamefont{B.~A.}
  \bibnamefont{Bernevig}}, \bibinfo{journal}{{Phys. Rev. B}}
  \textbf{\bibinfo{volume}{{80}}}, \bibinfo{pages}{180505}
  (\bibinfo{year}{{2009}}).

\bibitem[{\citenamefont{Cvetkovic and Tesanovic}({2009})}]{Cvetkovic09}
\bibinfo{author}{\bibfnamefont{V.}~\bibnamefont{Cvetkovic}} \bibnamefont{and}
  \bibinfo{author}{\bibfnamefont{Z.}~\bibnamefont{Tesanovic}},
  \bibinfo{journal}{{Europhys. Lett.}} \textbf{\bibinfo{volume}{{85}}},
  \bibinfo{pages}{37002} (\bibinfo{year}{{2009}}).

\bibitem[{\citenamefont{Stanev et~al.}({2008})\citenamefont{Stanev, Kang, and
  Tesanovic}}]{Stanev08}
\bibinfo{author}{\bibfnamefont{V.}~\bibnamefont{Stanev}},
  \bibinfo{author}{\bibfnamefont{J.}~\bibnamefont{Kang}}, \bibnamefont{and}
  \bibinfo{author}{\bibfnamefont{Z.}~\bibnamefont{Tesanovic}},
  \bibinfo{journal}{Phys. Rev. B} \textbf{\bibinfo{volume}{{78}}},
  \bibinfo{pages}{184509} (\bibinfo{year}{{2008}}).

\bibitem[{\citenamefont{Parker et~al.}({2009})\citenamefont{Parker, Vavilov,
  Chubukov, and Mazin}}]{parker}
\bibinfo{author}{\bibfnamefont{D.}~\bibnamefont{Parker}},
  \bibinfo{author}{\bibfnamefont{M.~G.} \bibnamefont{Vavilov}},
  \bibinfo{author}{\bibfnamefont{A.~V.} \bibnamefont{Chubukov}},
  \bibnamefont{and} \bibinfo{author}{\bibfnamefont{I.~I.} \bibnamefont{Mazin}},
  \bibinfo{journal}{Phys. Rev. B} \textbf{\bibinfo{volume}{{80}}},
  \bibinfo{pages}{100508} (\bibinfo{year}{{2009}}).

\bibitem[{\citenamefont{Vorontsov et~al.}({2009})\citenamefont{Vorontsov,
  Vavilov, and Chubukov}}]{anton}
\bibinfo{author}{\bibfnamefont{A.~B.} \bibnamefont{Vorontsov}},
  \bibinfo{author}{\bibfnamefont{M.~G.} \bibnamefont{Vavilov}},
  \bibnamefont{and} \bibinfo{author}{\bibfnamefont{A.~V.}
  \bibnamefont{Chubukov}}, \bibinfo{journal}{{Phys. Rev. B}}
  \textbf{\bibinfo{volume}{{79}}}, \bibinfo{eid}{{060508}}
  (\bibinfo{year}{{2009}}).

\bibitem[{\citenamefont{de~la Cruz et~al.}({2008})\citenamefont{de~la Cruz,
  Huang, Lynn, Li, II, Zarestky, Mook, Chen, Luo, Wang et~al.}}]{de08}
\bibinfo{author}{\bibfnamefont{C.}~\bibnamefont{de~la Cruz}},
  \bibinfo{author}{\bibfnamefont{Q.}~\bibnamefont{Huang}},
  \bibinfo{author}{\bibfnamefont{J.~W.} \bibnamefont{Lynn}},
  \bibinfo{author}{\bibfnamefont{J.}~\bibnamefont{Li}},
  \bibinfo{author}{\bibfnamefont{W.~R.} \bibnamefont{II}},
  \bibinfo{author}{\bibfnamefont{J.~L.} \bibnamefont{Zarestky}},
  \bibinfo{author}{\bibfnamefont{H.~A.} \bibnamefont{Mook}},
  \bibinfo{author}{\bibfnamefont{G.~F.} \bibnamefont{Chen}},
  \bibinfo{author}{\bibfnamefont{J.~L.} \bibnamefont{Luo}},
  \bibinfo{author}{\bibfnamefont{N.~L.} \bibnamefont{Wang}},
  \bibnamefont{et~al.}, \bibinfo{journal}{{Nature}}
  \textbf{\bibinfo{volume}{{453}}}, \bibinfo{pages}{899}
  (\bibinfo{year}{{2008}}).

\bibitem[{\citenamefont{Klauss et~al.}({2008})\citenamefont{Klauss, Luetkens,
  Klingeler, Hess, Litterst, Kraken, Korshunov, Eremin, Drechsler, Khasanov
  et~al.}}]{Klauss08}
\bibinfo{author}{\bibfnamefont{H.-H.} \bibnamefont{Klauss}},
  \bibinfo{author}{\bibfnamefont{H.}~\bibnamefont{Luetkens}},
  \bibinfo{author}{\bibfnamefont{R.}~\bibnamefont{Klingeler}},
  \bibinfo{author}{\bibfnamefont{C.}~\bibnamefont{Hess}},
  \bibinfo{author}{\bibfnamefont{F.~J.} \bibnamefont{Litterst}},
  \bibinfo{author}{\bibfnamefont{M.}~\bibnamefont{Kraken}},
  \bibinfo{author}{\bibfnamefont{M.~M.} \bibnamefont{Korshunov}},
  \bibinfo{author}{\bibfnamefont{I.}~\bibnamefont{Eremin}},
  \bibinfo{author}{\bibfnamefont{S.-L.} \bibnamefont{Drechsler}},
  \bibinfo{author}{\bibfnamefont{R.}~\bibnamefont{Khasanov}},
  \bibnamefont{et~al.}, \bibinfo{journal}{Phys. Rev. Lett.}
  \textbf{\bibinfo{volume}{{101}}}, \bibinfo{pages}{077005}
  (\bibinfo{year}{{2008}}).

\bibitem[{\citenamefont{Wang et~al.}({2009})\citenamefont{Wang, Zhai, Ran,
  Vishwanath, and Lee}}]{dhlee-PRL102}
\bibinfo{author}{\bibfnamefont{F.}~\bibnamefont{Wang}},
  \bibinfo{author}{\bibfnamefont{H.}~\bibnamefont{Zhai}},
  \bibinfo{author}{\bibfnamefont{Y.}~\bibnamefont{Ran}},
  \bibinfo{author}{\bibfnamefont{A.}~\bibnamefont{Vishwanath}},
  \bibnamefont{and} \bibinfo{author}{\bibfnamefont{D.-H.} \bibnamefont{Lee}},
  \bibinfo{journal}{Phys. Rev. Lett.} \textbf{\bibinfo{volume}{{102}}},
  \bibinfo{pages}{047005} (\bibinfo{year}{{2009}}).

\bibitem[{\citenamefont{Platt et~al.}({2009})\citenamefont{Platt, Honerkamp,
  and Hanke}}]{Platt09}
\bibinfo{author}{\bibfnamefont{C.}~\bibnamefont{Platt}},
  \bibinfo{author}{\bibfnamefont{C.}~\bibnamefont{Honerkamp}},
  \bibnamefont{and} \bibinfo{author}{\bibfnamefont{W.}~\bibnamefont{Hanke}},
  \bibinfo{journal}{New J. Phys.} \textbf{\bibinfo{volume}{{11}}},
  \bibinfo{pages}{055058} (\bibinfo{year}{{2009}}).

\bibitem[{\citenamefont{Lorenzana et~al.}({2008})\citenamefont{Lorenzana,
  Seibold, Ortix, and Grilli}}]{Lorenzana08}
\bibinfo{author}{\bibfnamefont{J.}~\bibnamefont{Lorenzana}},
  \bibinfo{author}{\bibfnamefont{G.}~\bibnamefont{Seibold}},
  \bibinfo{author}{\bibfnamefont{C.}~\bibnamefont{Ortix}}, \bibnamefont{and}
  \bibinfo{author}{\bibfnamefont{M.}~\bibnamefont{Grilli}},
  \bibinfo{journal}{Phys. Rev. Lett.} \textbf{\bibinfo{volume}{{101}}},
  \bibinfo{pages}{186402} (\bibinfo{year}{{2008}}).

\bibitem[{\citenamefont{Brydon and Timm}({2009})}]{Brydon09}
\bibinfo{author}{\bibfnamefont{P.}~\bibnamefont{Brydon}} \bibnamefont{and}
  \bibinfo{author}{\bibfnamefont{C.}~\bibnamefont{Timm}},
  \bibinfo{journal}{Phys. Rev. B} \textbf{\bibinfo{volume}{{79}}},
  \bibinfo{pages}{180504} (\bibinfo{year}{{2009}}).

\bibitem[{\citenamefont{Johannes and Mazin}({2009})}]{johannes:220510}
\bibinfo{author}{\bibfnamefont{M.~D.} \bibnamefont{Johannes}} \bibnamefont{and}
  \bibinfo{author}{\bibfnamefont{I.~I.} \bibnamefont{Mazin}},
  \bibinfo{journal}{{Phys. Rev. B}} \textbf{\bibinfo{volume}{{79}}}, \bibinfo{eid}{{220510}}
  (\bibinfo{year}{{2009}}).

\bibitem[{\citenamefont{Mazin and Johannes}({2009})}]{mazin09}
\bibinfo{author}{\bibfnamefont{I.~I.} \bibnamefont{Mazin}} \bibnamefont{and}
  \bibinfo{author}{\bibfnamefont{M.~D.} \bibnamefont{Johannes}},
  \bibinfo{journal}{{Nature Physics}} \textbf{\bibinfo{volume}{{5}}},
  \bibinfo{pages}{141} (\bibinfo{year}{{2009}}).

\bibitem[{\citenamefont{Eremin and Chubukov}({2009})}]{Eremin09}
\bibinfo{author}{\bibfnamefont{I.}~\bibnamefont{Eremin}} \bibnamefont{and}
  \bibinfo{author}{\bibfnamefont{A.~V.} \bibnamefont{Chubukov}},
  \bibinfo{howpublished}{arXiv:0911.1754v1} (\bibinfo{year}{{2009}}).

\end{thebibliography}
\end{document}